\documentclass[aps,preprint,showpacs,floatfix]{revtex4}
\usepackage{bm}
\begin{document}


\count255=\time\divide\count255 by 60 \xdef\hourmin{\number\count255}
  \multiply\count255 by-60\advance\count255 by\time
 \xdef\hourmin{\hourmin:\ifnum\count255<10 0\fi\the\count255}

\newcommand{\xbf}[1]{\mbox{\boldmath $ #1 $}}

\newcommand{\sixj}[6]{\mbox{$\left\{ \begin{array}{ccc} {#1} & {#2} &
{#3} \\ {#4} & {#5} & {#6} \end{array} \right\}$}}

\newcommand{\threej}[6]{\mbox{$\left( \begin{array}{ccc} {#1} & {#2} &
{#3} \\ {#4} & {#5} & {#6} \end{array} \right)$}}

\newcommand{\clebsch}[6]{\mbox{$\left( \begin{array}{cc|c} {#1} & {#2} &
{#3} \\ {#4} & {#5} & {#6} \end{array} \right)$}}

\newcommand{\iso}[6]{\mbox{$\left( \begin{array}{cc||c} {#1} & {#2} &
{#3} \\ {#4} & {#5} & {#6} \end{array} \right)$}}

\title{Phenomenology of the Baryon Resonance {\bf 70}-plet at Large $N_c$}

\author{Thomas D. Cohen}
\email{cohen@physics.umd.edu}

\affiliation{Department of Physics, University of Maryland, College
Park, MD 20742-4111}

\author{Richard F. Lebed}
\email{Richard.Lebed@asu.edu}

\affiliation{Department of Physics and Astronomy, Arizona State
University, Tempe, AZ 85287-1504}

\date{July, 2005}

\begin{abstract}
We examine the multiplet structure and decay channels of baryon
resonances in the large $N_c$ QCD generalization of the $N_c \! = \!
3$ SU(6) spin-flavor {\bf 70}.  We show that this ``{\bf 70}'',
while a construct of large $N_c$ quark models, actually consists of
five model-independent irreducible spin-flavor multiplets in the large
$N_c$ limit.  The preferred decay modes for these resonances
fundamentally depend upon which of the five multiplets to which the
resonance belongs.  For example, there exists an SU(3) ``{\bf 8}'' of
resonances that is $\eta$-philic and $\pi$-phobic, and an ``{\bf 8}''
that is the reverse.  Moreover, resonances with a strong SU(3) ``{\bf
1}'' component prefer to decay via a $\overline{K}$ rather than via a
$\pi$.  Remarkably, available data appears to bear out these
conclusions.
\end{abstract}

\pacs{11.15.Pg, 14.20.Gk, 14.20.Jn}

\maketitle

\section{Introduction} \label{intro}

Numerous scattering experiments performed during the past several
decades have generated a plethora of data revealing the excitation
behavior of baryons.  The most striking feature of this data at lower
energies is the existence of observable resonant states: the excited
baryons.  Inasmuch as QCD is the underlying fundamental theory of
strong interactions, the entire data set including the resonances
should be obtainable directly from QCD\@.  However, despite
considerable recent progress in the treatment of excited states in
lattice QCD~\cite{lattice}, the extraction of resonant state
properties {\it ab initio\/} from QCD remains a very hard problem.
Indeed, first-principles QCD has so far yielded no simple explanation
for the mere existence of resonances narrow enough to be resolved.
Thus, to a very large extent most of our insight into these resonant
states is gleaned from models, such as the constituent quark model,
whose connection to full QCD remains obscure.  Given this
unsatisfactory situation, it is useful to ask whether there are any
known systematic approaches to QCD that can give some qualitative or
semi-quantitative insight into aspects of baryon resonances,
independent of models.  In a series of
papers~\cite{CL1st,CLcompat,CLconfig,CL1N,CLphoto,CLpenta,CLSU3,CLSU3penta}
(see Ref.~\cite{review} for short reviews), we have argued that large
$N_c$ QCD and the $1/N_c$ expansion about this limit provide just such
an approach.  In this paper we explore the formal and phenomenological
implications of this approach for the states which, in the
conventional quark model language, are collected into an SU(6)
70-plet.

An important caveat is necessary at the outset: In a hypothetical
world where $N_c$ is truly large, the $1/N_c$ expansion is clearly
valid and provides very accurate predictions.  However, in the real
world $N_c \! = \! 3$, and $1/N_c$ corrections can be substantial.
While for some observables ({\it e.g.}, masses of stable baryons) the
leading-order large $N_c$ predictions give reasonable qualitative and
often semi-quantitative descriptions of the world, for others ({\it
e.g.}, scalar meson properties) the large $N_c$ predictions are quite
poor.  Thus the question of whether large $N_c$ analysis is merely an
exercise in mathematical physics or a useful phenomenological tool
depends on which observables are being studied.  While a number of
interesting phenomenological predictions of baryon resonance
observables have already been obtained from such analyses, the exact
extent to which the approach successfully describes this sector
remains an open question.  In part, this paper addresses the issue by
showing how certain qualitative features observed in the decays of
70-plet states can be understood in the context of large $N_c$ QCD.

Two principal ideas underlie the model-independent large $N_c$
approach to excited baryons developed in
Refs.~\cite{CL1st,CLcompat,CLconfig,CL1N,CLphoto,CLpenta,CLSU3,CLSU3penta}.
First, one must focus from the outset directly on the physical
scattering observables from which resonances are ultimately extracted
(meson-nucleon, electroproduction, or photoproduction scattering
amplitudes) rather than on the resonance positions themselves.
Second, such scattering amplitudes can be represented as operators to
be evaluated between asymptotic meson-baryon states, and as such are
subject to the contracted SU(2$N_f$) symmetry ($N_f$ being the number
of light quark flavors) known to emerge from a model-independent
analysis based upon large $N_c$ consistency relations~\cite{DJM1}.
Combining these two ideas allows one to derive expressions, true at
large $N_c$, for the scattering amplitude in any given channel to be
expressed as a sum of terms consisting of group-theoretical factors
multiplied by {\em reduced amplitudes}.  As there are fewer reduced
amplitudes than observable scattering amplitudes, the approach has
predictive power: the scattering amplitudes in different channels are
related at large $N_c$, and one predicts that various linear
combinations of amplitudes are equal at large $N_c$.

It should be noted that this analysis does not by itself predict the
existence of any baryon resonances {\it ab initio}.  Generic large
$N_c$ counting gives excited baryon widths of $O(N_c^0)$, which is the
same order as the spacing between states.  Whether at large $N_c$ such
states are sufficiently narrow to resolve is a matter of dynamical
detail and not generic large $N_c$ scaling.  The spacing of baryon
resonances differs sharply from that of excited mesons, which have
widths of $O(1/N_c)$ and hence are narrow at large $N_c$.  However,
this analysis does make a definitive prediction about resonances: Any
resonances that do exist must fall into multiplets that become
degenerate in both mass and width (or equivalently, coupling constant)
at large $N_c$.  The reason is simple: Resonances are poles in the
scattering amplitudes, and at large $N_c$ these amplitudes are
entirely determined by the reduced amplitudes.  Hence, a resonance in
some channel implies a pole in a reduced amplitude.  However, each
reduced amplitude contributes to multiple physical amplitudes, each of
which therefore has a resonance at the same location [up to $O(1/N_c)$
corrections].  The pattern of the degeneracy is fully fixed by the
contracted SU(2$N_f$) symmetry.  For $N_f \! = \! 2$ these degenerate
multiplets are completely determined by a single quantum number $K$
that emerges from the analysis.

The scheme outlined above is fully model independent and exact at
large $N_c$.  If one makes a further assumption about
resonances---namely, that decay channels near a resonant energy are
dominated by the resonance rather than the continuum---then one can
also use the contracted SU(2$N_f$) symmetry to deduce selection rules
for the decays that hold at large $N_c$.  This additional assumption
is needed since the extraction of resonance branching ratios (BR) from
scattering data is intrinsically model dependent.  However, to the
extent that the amplitude is dominated by the resonance (in a limited
kinematical region), this model dependence becomes small.  As noted
above, large $N_c$ analysis alone does not imply that baryon
resonances even exist and clearly gives no guidance on the question of
whether the resonances are sufficiently prominent for the meaningful
extraction of BR\@.  In this work we rely on the phenomenological fact
that prominent resonances are known to exist in the region of
interest, 1.4--2.0~GeV.

Much of the early work based on this
approach~\cite{CL1st,CLcompat,CLconfig,CL1N,CLphoto,CLpenta} was
limited to nonstrange particles.  While the intellectual underpinnings
are the same regardless of $N_f$, the inclusion of strange quarks
complicates the analysis in important technical ways, particularly in
the limit of SU(3) flavor symmetry.  The large $N_c$ analogues to the
2-flavor physical states possess isospin quantum numbers identical
to those at $N_c \! = \! 3$.  Additional representations also arise,
but these are dismissed as ``large $N_c$ artifacts''.  However, in the
case of three degenerate flavors, {\em none\/} of the $N_c \! = \! 3$
SU(3) flavor representations for baryons remain the same dimension as
their $N_c \!  > \! 3$ generalizations~\cite{DJM1,DP}.

In fact, {\em all\/} of the large $N_c$ baryon multiplets are infinite
dimensional as $N_c \! \to \! \infty$ [see Eq.~(\ref{dimsize})].  This
raises two issues when one attempts to relate large $N_c$ predictions
to the real world.  The first is how to associate a given large $N_c$
multiplet with an $N_c \! = \! 3$ multiplet.  The second is that,
since large $N_c$ multiplets are (infinitely) larger than their $N_c
\! = \! 3$ cousins, one needs to prescribe how to associate states
in the analogous $N_c \! = \! 3$ representation with states in the
large $N_c$ representation.  The first issue is easily resolved: There
is an obvious association between $N_c \! = \! 3$ representations with
representations for any arbitrary $N_c$, which is given explicitly in
the following section.  One then computes quantities at arbitrary
$N_c$ and takes the large $N_c$ limit.  To make manifest the
connection between the large $N_c$ and $N_c \! = \!  3$
representations, we adopt the convention of denoting the large $N_c$
analogue of the baryon {\bf 8} as ``{\bf 8}'', the analogue of the
{\bf 10} as ``{\bf 10}'', and so forth.  The second issue is also
relatively straightforward to resolve: One considers only those states
within a multiplet with the same values of isospin and strangeness as
occur for $N_c \! = \! 3$.

Since the baryon representations increase in dimension with $N_c$, the
SU(3) Clebsch-Gordan coefficients (CGC) needed for this analysis are
not tabulated in standard sources.  Instead, one requires the
$N_c$-dependent CGC computed and tabulated in Ref.~\cite{CLSU3}.  As
discussed in this paper, the CGC implicitly impose formally and
phenomenologically interesting selection rules as $N_c$ becomes large.

The basic analysis of
Refs.~\cite{CL1st,CLcompat,CLconfig,CL1N,CLphoto,CLpenta,CLSU3,CLSU3penta}
is fully model independent.  Another approach to excited baryons at
large $N_c$ uses large $N_c$ generalizations of the quark model, or at
least a quark ``picture'' in which the quantum numbers of the $N_c$
quarks are the important degrees of
freedom~\cite{NStar_old,nonstrange,strange}.  The large $N_c$ quark
model has the same emergent symmetries as large $N_c$ QCD\@.  Thus, if
one focuses entirely on those properties that are related to the
symmetry, the large $N_c$ quark model may be viewed as an efficient
way to deduce group-theoretical results. It was shown explicitly in
the case of the mixed-symmetry (MS) $N_f \! = \! 2$ ``{\bf 20}''-plet
of SU(2$N_f$) associated with $\ell \! = \! 1$ orbital excitations
that the patterns of degeneracy from the large $N_c$ quark model are
compatible with the degeneracy patterns among resonances directly
deduced from large $N_c$ QCD~\cite{CLcompat}.  One of the purposes of
the present paper is to show explicitly that the same compatibility
holds for the $N_f \! = \! 3$ MS ``{\bf 70}''-plet states of the large
$N_c$ quark model: At leading ($N_c^0$) order the states fall into
multiplets which are compatible with the degeneracy patterns deduced
from full large $N_c$ QCD.

The technical advantages of the method based on the large $N_c$ quark
model are quite apparent: It is elegant and efficient to classify
quark model operators in terms of their $N_c$ scaling behavior.  Since
many operators connecting states are subleading in $1/N_c$ counting,
the approach constrains the possible eigenstates, which in turn
generates degenerate multiplets at large $N_c$.

It is worth noting, however, that the large $N_c$ quark model builds
in dynamics beyond the emergent symmetry.  All of this dynamics is
model dependent and thus cannot be taken as direct predictions of
large $N_c$ QCD\@.  The model-dependent aspects include: $i$) the
existence of the resonances, $ii$) the fact that the resonances have
negligible widths ({\it i.e.}, are stable) in the model, with widths
only added in via an {\it ad hoc\/} prescription, and $iii$)
assumptions about the detailed nature of the state.  The third aspect
is particularly important: Models used typically assume that the
states fall into unmixed configurations of SU($2N_f$)$\times$O(3); the
physical picture behind this symmetry is that there is only a single
orbitally excited quark [giving rise to the O(3)] on top of a
spherically symmetric core with an SU(2$N_f$) spin-flavor symmetry.
This assumption does not follow from large $N_c$ QCD\@.  As noted in
Ref.~\cite{CLconfig}, configuration mixing of states of this type can
occur at leading ($N_c^0$) order.  In large $N_c$ QCD the only
emergent symmetry is SU(2$N_f$), which refers to the entire state and
is {\em not\/} a symmetry of the spin and flavor of individual quarks,
as is the case for the excited states in the unmixed quark model.
Thus in large $N_c$ QCD it is not meaningful to ask whether the
spin-flavor symmetry is in a pure MS state such as the ``{\bf
70}''-plet of the quark model.

Given these problems, one might simply avoid using quark model
language entirely and rely exclusively on the symmetries of large
$N_c$ QCD\@. While we generally advocate this view, it is useful
nonetheless to make contact with the quark model picture since this
picture informs so much of our intuition about excited
baryons. Accordingly, in previous papers~\cite{CL1st,CLcompat} we
identified the states in the excited SU(4) ``{\bf 20}''-plet in terms
of complete multiplets labeled by the $K$ quantum number.  In this
paper we generalize the analysis to three flavors and extend the
analysis to the SU(6) ``{\bf 70}''-plet. In particular, we find that
the (``{\bf 70}''$\!$,$\, 1^-$) of SU(6)$\times$O(3) is a reducible
multiplet in large $N_c$, consisting [in the SU(3) limit] of 5
complete multiplets labeled by $K$.

This paper has four main purposes.  The first is to flesh out the
3-flavor version of the model-independent approach that was briefly
described in Ref.~\cite{CLSU3penta}; the second is to point out the
existence of SU(3)-flavor selection rules that emerge at large $N_c$.
The third is to tie the general scattering approach to the quark
model-based approach for the ``{\bf 70}''-plet states; and the fourth
is to apply these methods to describe phenomenologically the decays of
the {\bf 70}-plet states (or more precisely, the states that are
typically assigned to the {\bf 70}-plet in quark models).

This paper is organized as follows: In Sec.~\ref{group} we provide
essential group-theoretical background and establish notation.
Section~\ref{CGCthm} presents a salient property of SU(3)
Clebsch-Gordan coefficients at large $N_c$ that is useful in obtaining
information about processes involving strange resonances.
Section~\ref{3to2} shows the explicit connection between 3-flavor and
2-flavor scattering amplitude expressions.  In Sec.~\ref{5mult} we
show that the ``{\bf 70}'' consists of 5 multiplets labeled by $K$,
and exhibit the connection to quark-picture operators.
Section~\ref{phenom} provides a number of phenomenological
consequences of our results, and Sec.~\ref{concl} concludes.

\section{Group Theory Preliminaries} \label{group}

Much of the content of this Section appears in Ref.~\cite{CLSU3} and
is presented here for the reader's convenience.  An irreducible
representation ({\it irrep}) of SU(3) symmetry may be denoted by its
Dynkin weights $(p,q)$, which indicate a Young tableau with $p \! + \!
q$ boxes in the top row and $q$ boxes in the bottom row.  In terms of
the maximal value of hypercharge and the isospin of the
singly-degenerate states of the top row in the SU(3) weight diagram,
one finds $Y_{\rm max} \! = \! \frac 1 3 (p \! + \! 2q)$ and $I_{\rm
top} \! = \! \frac 1 2 p$.

Mesons at arbitrary $N_c$ still carry the quantum numbers of a single
$q\bar q$ pair, and hence their SU(3) flavor irreps are unchanged when
$N_c$ is changed.  The SU(3) irreps may also be denoted as usual by
their dimensions, if no ambiguity arises: {\it e.g.}, {\bf 8} = (1,1).

Baryons, on the other hand, carry the quantum numbers of $N_c$ quarks
[in order to form an SU($N_c$) color singlet from color-fundamental
irreps], and therefore the baryon SU(3) flavor irreps grow in size
with $N_c$.  The baryon SU(3) irreps $R$ corresponding to large $N_c$
generalizations of those occurring at $N_c \! = \! 3$ are taken to be
$R \! = \!  (2I_{\rm top}, \frac{N_c}{2} + \frac {3r}{2} - I_{\rm
top})$, which has $Y_{\rm max} \! = \! \frac{N_c}{3} + r$.  The
quantity $r \! = \!  O(N_c^0)$ is an integer, as required by
quantization of the Wess-Zumino term for arbitrary $N_c$~\cite{WZ}.
The dimension $[R] \!  = \! \frac 1 2 (p+1)(q+1)(p+q+2)$ of an
arbitrary SU(3) irrep assumes a useful limiting expression for the
large $N_c$ baryon irreps:
\begin{equation} \label{dimsize}
[R] \longrightarrow \frac{N_c^2}{8} [I_{\rm top}] \ {\rm as} \ N_c
\to \infty \ ,
\end{equation}
where $[I_{\rm top}] \! = \! 2I_{\rm top} \! + 1$ is the isomultiplet
dimension.  A baryon irrep that generalizes a familiar $N_c \! = \! 3$
counterpart may also be denoted by its $N_c
\! = \! 3$ dimension within quote marks; the ones useful to this work
are
\begin{eqnarray}
\mbox{\rm ``{\bf 1}''} \equiv [ \, 0, (N_c \! - \! 3)/2 \, ] \ , & &
\mbox{\rm ``{\bf 8}''} \equiv [ \, 1, (N_c \! - \! 1)/2 \, ] \ ,
\nonumber \\
\mbox{\rm ``{\bf 10}''} \equiv [ \, 3, (N_c \! - \! 3)/2 \, ] \ , & &
``\overline{\bf 10}\mbox{''} \equiv [ \, 0, (N_c \! + \! 3)/2 \,
] \ ,
\nonumber \\
\mbox{\rm ``{\bf 27}''} \equiv [ \, 2, (N_c \! + \! 1)/2 \, ] \ , & &
\mbox{\rm ``{\bf 35}''} \equiv [ \, 4, (N_c \! - \! 1)/2 \, ] \ .
\end{eqnarray}
Other irreps appear only for $N_c \! > \!  3$ and are denoted only by
their Dynkin weights.  An exception that is useful to us in the
following is
\begin{equation}
\mbox{\rm ``{\bf S}''} \equiv [2, (N_c \! - \! 5)/2] \ ,
\end{equation}
so named because its $Y_{\rm max} \! = \frac{N_c}{3} \! - \! 1$
isomultiplet has $I \! = \! 1$, {\it i.e.}, $\Sigma$ quantum numbers.

SU(3) CGC are indicated by the notation
\begin{equation}
\clebsch{R_1}{R_2}{R_\gamma}{I_1,I_{1z},Y_1}{I_2,I_{2z},Y_2}{I,I_z,Y}
= \iso{R_1}{R_2}{R_\gamma}{I_1,Y_1}{I_2,Y_2}{I,Y}
\clebsch{I_1}{I_2}{I}{I_{1z}}{I_{2z}}{I_z} \ ,
\end{equation}
where the last factor is an ordinary SU(2) isospin CGC\@.  The
quantities containing a double bar, which do not depend upon the
additive $I_z$ quantum numbers, are called SU(3) isoscalar factors
and, like the full CGC, form orthogonal matrices.  We may refer to the
SU(3) isoscalar factors themselves as CGC if no ambiguity arises.  The
subscript $\gamma$ indicates possible distinct copies of a particular
irrep $R$ within the product $R_1 \! \otimes \! R_2$.

The SU(3) products phenomenologically useful in meson-baryon
scattering are
\begin{eqnarray}
\mbox{\rm ``{\bf 8}''} \otimes {\bf 8} & = & \mbox{\rm ``{\bf 27}''}
\oplus \mbox{\rm ``{\bf 10}''} \oplus ``\overline{\bf 10}\mbox{\rm ''}
\oplus ``{\bf 8}_1 \mbox{\rm ''}
\oplus ``{\bf 8}_2 \mbox{\rm ''} \oplus ``{\bf 1} \mbox{\rm ''} \oplus
``{\bf S} \mbox{\rm ''} \ , \nonumber \\
\mbox{\rm ``{\bf 10}''} \otimes {\bf 8} & = & \mbox{\rm ``{\bf 35}''}
\oplus \mbox{\rm ``{\bf 27}''} \oplus ``{\bf 10}_1 \mbox{\rm ''}
\oplus ``{\bf 10}_2 \mbox{\rm ''} \oplus ``{\bf 8} \mbox{\rm ''}
\oplus ``{\bf S} \mbox{\rm ''} \! \oplus
[ \, 5,(N_c \! - \! 5)/2 \, ] \oplus [ \, 4,(N_c \! - \! 7)/2 \, ] .
\nonumber \\ & & \label{prods}
\end{eqnarray}
The final two irreps need not be considered further, not only because
they are absent for $N_c \! = \! 3$, but also (unlike ``{\bf S}'') do
not contain any isomultiplets with $N_c \! = \! 3$ quantum numbers.
In Ref.~\cite{CLSU3}, ${\bf 10}_1$ is defined as the unique product
{\bf 10} irrep whose ``{\bf 10}''$\otimes \,${\bf 8} CGC all vanish
with powers of $N_c \! - \! 3$ (which occurs because only one {\bf 10}
appears in the $N_c \! = \! 3$ product {\bf 10}$\, \otimes
\,${\bf 8}).  The CGC for ${\bf 10}_1$ and ``{\bf S}'' are not needed
for strict $N_c \! = \! 3$ phenomenology and therefore were not
compiled in Ref.~\cite{CLSU3}, but are useful for formal large $N_c$
results requiring unitarity at arbitrary $N_c$, as is employed in the
following analysis.

A similar notation may also be extended to SU(6) spin-flavor
multiplets.  As shown long ago, the requirement of order-by-order
unitarity in powers of $1/N_c$ in meson-baryon scattering requires
that the $J^P \!  = \! \frac{1}{2}^+$ ``{\bf 8}'' and the $J^P \! = \!
\frac{3}{2}^+$ ``{\bf 10}'' belong to a single spin-flavor
multiplet~\cite{DJM1} whose members differ in mass only at
$O(1/N_c)$~\cite{Jenk,DJM2}, the completely symmetric SU(6) ``{\bf
56}'' $\equiv \! (N_c, 0, 0, 0, 0, 0)$.  For $N_c \! > \! 3$ the
``{\bf 56}'' also contains $J^P \! = \! \frac{5}{2}^+ \! , \ldots ,
\frac{N_c}{2}^+$ flavor multiplets.   Since the physical $N_c \! = \!
3$ {\bf 8} baryons are stable against strong decay, the same is true
for the full ``{\bf 56}'' when $N_c$ is sufficiently large; hence the
``{\bf 56}'' is labeled the ``ground-state'' band.

In the SU(6) quark model, the first excited multiplet consists of
states symmetric on all except one of the quarks, the {\bf 70}-plet.
The exceptional quark is then combined with the symmetric ``core'' as
an $\ell \! = \!  1$ orbital excitation.  We denote the analogue for
arbitrary $N_c$ as ``{\bf 70}'' $\equiv \! (N_c \! - \!  2, 1, 0, 0,
0)$.  Its decomposition into SU(3)$\times$SU(2) [the total spin SU(2)
factor including not only quark spin but the orbital angular momentum
as well] gives numerous spin-flavor multiplets~\cite{CLcompat}, but
only those multiplets containing states with $N$, $\Delta$, $\Lambda$,
$\Sigma$, $\Xi$, and $\Omega$ quantum numbers and spins possible with
3 quarks are phenomenologically relevant.  These multiplets are
\begin{eqnarray} &&
2 \left( ``{\bf 8}\mbox{''}, \frac 1 2 \right) \!
\oplus 2 \left( ``{\bf 8}\mbox{''}, \frac 3 2 \right) \!
\oplus \left( ``{\bf 8}\mbox{''}, \frac 5 2 \right) \!
\oplus 2[1] \left( ``{\bf 10}\mbox{''}, \frac 1 2 \right) \!
\oplus 3[1] \left( ``{\bf 10}\mbox{''}, \frac 3 2 \right) \!
\oplus 2[0] \left( ``{\bf 10}\mbox{''}, \frac 5 2 \right) \!
\oplus
\nonumber \\ &&
1[0] \left( ``{\bf 10}\mbox{''}, \frac 7 2 \right) \!
\oplus \left( ``{\bf 1}\mbox{''} , \frac 1 2 \right) \!
\oplus \left( ``{\bf 1}\mbox{''} , \frac 3 2 \right) \!
\oplus 2[0] \left( ``{\bf S}\mbox{''}, \frac 1 2 \right) \!
\oplus 2[0] \left( ``{\bf S}\mbox{''}, \frac 3 2 \right) \!
\oplus 1[0] \left( ``{\bf S}\mbox{''}, \frac 5 2 \right)
\label{70mult}
\end{eqnarray}
where the coefficients indicate multiplicities for $N_c$ large (and
for $N_c \! = \! 3$ in brackets, if different).  For $N_c \! > \! 3$
the ``{\bf 8}'' and ``{\bf 10}'' contain no additional states with
$N_c \! = \! 3$ quantum numbers, but ``{\bf 1}'' gains a $\Xi$, and
``{\bf S}'' has $\Sigma$, $\Xi$, and $\Omega$ states.  One of our
results below is that in the absence of SU(3) breaking the 20
[SU(3),$\,$SU(2)] multiplets have only 5 distinct masses split at
$O(N_c^0)$, meaning that the large $N_c$ SU(6)$\times$O(3) (``{\bf
70}''$\!$,$\, 1^-$) is actually reducible to 5 distinct multiplets.

\section{A Property of SU(3) CGC} \label{CGCthm}

Much of the power of the analysis rests on an observation that holds
for all arbitrary-$N_c$ SU(3) CGC thus far computed, which includes
every coupling relevant to $N_c \! = \! 3$ phenomenology.  We do not
prove this result exhaustively as a theorem, but rather show below by
direct mathematical construction how it arises.  But first, we state
the property:

Let $R_B \! = \! (2S_B, \, \frac{N_c}{2} \! - \! S_B)$ denote an SU(3)
irrep [corresponding to baryons in the ground-state SU(6) ``{\bf 56}''
with spin $S_B$, for which the top (nonstrange) row in the weight
diagram has isospin $I_{B, \rm top} \!  = \! S_B$ and $Y_{B, \rm max}
\! = \! \frac{N_c}{3}$], let $R_\phi \!  = \!  (p_\phi, \, q_\phi)$ be
an SU(3) (meson) irrep with weights $p_\phi$, $q_\phi \!  = \!
O(N_c^0)$, and let $R_s \gamma_s \subset R_B \otimes R_\phi$, where
$Y_{s, \rm max} \! = \! \frac{N_c}{3} + r$ and $R_s = (2I_{s, \rm
top}, \, \frac{N_c}{2} + \frac{3r}{2} - I_{s, \rm top})$, $r \! = \!
O(N_c^0)$.  Then the SU(3) CGC satisfy
\begin{equation} \label{CGCmag}
\left( \begin{array}{cc||c}
R_B & R_\phi & R_s \, \gamma_s \\ I_B, \frac{N_c}{3} \! - \! m &
I_\phi, Y_\phi & I_s, \, \frac{N_c}{3} \! + \! Y_\phi \! - \! m
\end{array} \right) \le O(N_c^{-|Y_\phi - r|/2}) \ ,
\end{equation}
for all allowed $O(N_c^0)$ values of $m$, saturation of the inequality
occurring for almost all CGC\@.  One may observe this remarkable fact
in the tables of Ref.~\cite{CLSU3}.

This interesting property indicates that baryon resonances in various
SU(3) irreps preferentially couple to mesons with a unique value of
hypercharge.  In particular, those with $Y_{\rm max} \! = \!
\frac{N_c}{3} + 1$ (``$\overline{\bf 10}$'', ``{\bf 27}'', and
``{\bf 35}'') decay via a $K^+$ or $K^0$, those with $Y_{\rm max} \! =
\! \frac{N_c}{3}$ (``{\bf 8}'' and ``{\bf 10}'') decay via $\pi$ or
$\eta$, and those with $Y_{\rm max} \! = \! \frac{N_c}{3} \! - \! 1$
(``{\bf 1}'' and ``{\bf S}'') decay via $\overline{K^0}$ or $K^-$.

The property Eq.~(\ref{CGCmag}) results from a combination of
unitarity and completeness of the SU(3) CGC, in addition to the
$U$-spin and $V$-spin values of the states in question.  Unitarity and
completeness require that, for every choice of $R_B$, $I_B$, $Y_B$ and
$R_\phi$, $I_\phi$, $Y_\phi$ with total coupled isospin $I_s$ and
hypercharge $Y_s$, there must exist at least one product irrep $R_s
\gamma_s$ whose corresponding CGC assumes the largest allowed
magnitude, $O(N_c^0)$.  One may therefore begin with ``stretched''
quantum numbers, for which precisely one $R_s$ is allowed and the
corresponding CGC are therefore guaranteed to be $O(N_c^0)$---indeed,
unity in magnitude.  For example, the state in the product ``{\bf
8}''$\otimes \,${\bf 8}$\, \subset$``$\overline{\bf 10}$'' with $I_s
\!  = \! 0$, $Y \! = \! \frac{N_c}{3} \! + \! 1$ is the only one in
the product of $I_B \! = \! \frac 1 2$, $Y_B \! = \!
\frac{N_c}{3}$ and $I_\phi \! = \! \frac 1 2$, $Y_\phi \! = \! 1$, and
therefore its CGC is $O(N_c^0)$ ($-1$, in fact).

Now note that the $U_\pm$ and $V_\pm$ SU(3) ladder operators assume a
very useful form~\cite{Bied}.  For example,
\begin{eqnarray}
V_+ | \, (p,q) \, I I_z Y \rangle & = & +g[(p,q), \, I, I_z, Y] \ |
\, (p,q), \, I \! + \! 1/2, \, I_z \! + \! 1/2, \, Y \! \! + \! 1 \,
\rangle \nonumber \\ & &
+g[(p,q), \, -(I \! + \! 1), \, I_z, \, Y] \ | \, (p,q), \, I \! - \!
1/2, \, I_z \! + \! 1/2, \, Y \! \! + \! 1 \, \rangle \ , \label{Vplus}
\end{eqnarray}
where the function $g$ is given by~\cite{CLSU3}
\begin{equation}
g[(p,q) \, I I_z Y] = \left\{ \frac{(I \! + \! I_z \! + \! 1)[\frac 1
3 (p \! - \! q) \! + \! I \! + \! \frac Y 2 \! + \! 1][\frac 1 3 (p \!
+ \! 2q) \! + \! I \! + \! \frac Y 2 \! + \! 2][\frac 1 3 (2p \! + \!
q) \! - \! I \! - \! \frac Y 2]}{(2I \! + \! 1)(2I \! + \! 2)}
\right\}^{1/2} ,
\end{equation}
and is the analogue to the familiar SU(2) functions $[(I \mp I_z) (I
\pm I_z +1)]^{1/2}$ that appear with the operators $I_\pm$.  SU(3)
CGC are then derived by the same coupling approach as for SU(2) ({\it
e.g.}, $V_{s,\pm} \!  = \! V_{B,\pm} \! + \! V_{\phi , \pm}$).  As
seen from Eq.~(\ref{Vplus}), these ladder operators generally produce
two states, and therefore the SU(3) recursion relations generally
involve six CGC~\cite{Bied,CLSU3}.  We decline to present these
cumbersome expressions here ({\it e.g.}, Eq.~(2.5) in~\cite{CLSU3}),
but merely indicate features important to the current analysis.

Since the meson irrep $R_\phi$ does not scale with $N_c$, the
functions $g$ appearing from $U_{\phi, \pm}$ or $V_{\phi, \pm}$, which
have $|\Delta Y_\phi| \! = \! 1$, do not induce any $N_c$ factors.
However, for the baryon irreps $R_B$ and $R_s$, the quantities $q$ and
$Y$ appearing in $U_{B, \pm}$, $V_{B, \pm}$, $U_{s, \pm}$, and $V_{s,
\pm}$ (all of which have $\Delta Y_\phi \! = \! 0$) both scale as
$N_c$.  Interestingly, two of the three factors in $g$ containing $q$
and $Y$ appear in the combination $\frac q 3 \! - \! \frac Y 2$, whose
$O(N_c)$ term cancels, while the $O(N_c)$ term in the third (in the
combination $\frac{2q}{3} \! + \! \frac{Y}{2}$) does not, making the
corresponding $g$ factors $O(N_c^{1/2})$.  This factor can also be
seen from the fact that the given states, lying near $Y_{\rm max}$,
are linear combinations of eigenstates carrying large values of $U$-
and $V$-spin and near-maximal values of $U_3$ and $V_3$.  Since the
$g$-factors are simply $[(U \mp U_3)(U \pm U_3 + 1)]^{1/2}$ and $[(V
\mp V_3)(V \pm V_3 + 1)]^{1/2}$ in disguise, each one has but a single
$O(N_c^{1/2})$ factor.  Dividing through by this $N_c^{1/2}$, the
$U_\pm$, $V_\pm$ CGC recursion relations assume the form
\begin{equation} \label{CGCorder}
{\rm (4 \ CGC \ with \ \Delta Y_\phi \! = \! 0)} +
\frac{1}{\sqrt{N_c}} {\rm (2 \ CGC \ with \ |\Delta Y_\phi| \! = \!
1)} = 0 \ .
\end{equation}
This result indicates that all CGC with $\Delta Y_\phi \! = \! 0$ tend
to appear at the same order in $N_c$, barring a fortuitous
cancellation.  However, since the 6-CGC recursion relations also
include ordinary SU(2) isospin CGC [again, see Eq.~(2.5)
in~\cite{CLSU3} for an example], and the same SU(3) CGC appear for
several independent charge states, such cancellations are
comparatively rare.

In practice, one begins with the stretched states for all $R_s
\gamma_s \! \subset \! R_B \! \otimes R_\phi$ that have the
largest allowed $Y_{s, \rm max}$ value, which therefore have
$O(N_c^0)$ CGC, and uses the $U_\pm$, $V_\pm$ recursion relations to
obtain all other $O(N_c^0)$ CGC for the given $R_s \gamma_s$, all of
which [by Eq.~(\ref{CGCorder})] have the same value of $Y_\phi \!  =
\!  Y_{s, \rm max} \! - \! Y_{B, \rm max}$, where $Y_{\rm B,
\rm max} \! = \! \frac{N_c}{3}$.  For example, in ``{\bf 8}''$\otimes
\,${\bf 8} the largest $Y_{s, \rm max}$ value $\frac{N_c}{3} + 1$ is
obtained for $R_s \! =$``{\bf 27}'' and ``$\overline{\bf 10}$'', and
the $O(N_c^0)$ CGC for these two product irreps all have $Y_\phi \! =
\! 1$.  Order-by-order unitarity in $N_c$ and the completeness of
SU(3) CGC with a given fixed value of $Y_\phi$ then imply that the
$O(N_c^0)$ CGC thus obtained are the only ones carrying the given
$Y_\phi$ value.  But then, Eq.~(\ref{CGCorder}) and unitarity imply
that changing the value of $Y_\phi$ by one unit (call it
$Y_{\phi^\prime}$) for the same $R_s \gamma_s$ produces CGC that are
generically a factor $N_c^{-1/2}$ smaller.  By completeness, other
irreps $R_{s^\prime} \gamma_{s^\prime}$ must step in to provide the
$O(N_c^0)$ CGC for the new value $Y_{\phi^\prime}$, and by noting
again that the stretched cases (those carrying $Y_{s^\prime, \rm
max}$) must have $O(N_c^0)$ CGC, we see from $Y_{\phi^\prime} \!  = \!
Y_{s^\prime, \rm max} - Y_{B, \rm max}$ that the value of
$Y_{s^\prime, \rm max}$ must also change by one unit.  In the ``{\bf
8}''$\otimes \,${\bf 8} example, the $Y_\phi \! = \! 0$ CGC with $R_s
\!  =$``{\bf 27}'' and ``$\overline{\bf 10}$'' are at most
$O(N_c^{-1/2})$, meaning that the remaining $R_s \gamma_s$ with $Y_{s,
\rm max} \! = \! \frac{N_c}{3}$ states (``${\bf 8}_\gamma$'' and {\bf
10}) must provide the $O(N_c^0)$ CGC\@.  Continuing the construction
in this fashion establishes Eq.~(\ref{CGCmag}).

\section{Reducing the 3-Flavor Case to the 2-Flavor Case} \label{3to2}

The master amplitude expression for a 3-flavor meson-baryon scattering
process $\phi (S_\phi, R_\phi, I_\phi, Y_\phi) + B (S_B, R_B, I_B,
Y_B) \to \phi^\prime (S_{\phi^\prime}, R_{\phi^\prime},
I_{\phi^\prime}, Y_{\phi^\prime}) + B^\prime (S_{B^\prime},
R_{B^\prime}, I_{B^\prime}, Y_{B^\prime})$, where $S$ is particle
spin, was originally obtained in Ref.~\cite{MMSU3}, and generalized to
include $O(N_c)$ quantum numbers in Ref.~\cite{CLSU3penta}.  The
master expression for such scattering amplitudes in the large $N_c$
limit then reads
\begin{eqnarray}
\lefteqn{S_{L L^\prime S S^\prime J_s R_s \gamma_s \gamma^\prime_s I_s
Y_s}} \nonumber \\
& = & (-1)^{S_B - S_{B^\prime}}
([R_B][R_{B^\prime}][S][S^\prime])^{1/2} / [R_s]
\sum_{\stackrel{\scriptstyle I \in R_\phi, \; I^\prime \in
R_{\phi^\prime},}{I^{\prime\prime} \in R_s, \; Y \in R_\phi \cap
R_{\phi^\prime}}} (-1)^{I + I^\prime + Y} [I^{\prime\prime}]
\nonumber \\ & & \times
\left( \begin{array}{cc||c} R_B & R_\phi & R_s \, \gamma_s \\ S_B
\frac{N_c}{3} & I Y & I^{\prime\prime} \, Y \! \! + \! \frac{N_c}{3}
\end{array} \right)
\left( \begin{array}{cc||c} R_B & R_\phi & R_s \, \gamma_s \\ I_B
Y_B & I_\phi Y_\phi & I_s Y_s \end{array} \right) \nonumber \\
& & \times
\left( \begin{array}{cc||c} R_{B^\prime} & R_{\phi^\prime} & R_s \,
\gamma^\prime_s \\ S_{B^\prime}
\frac{N_c}{3} & I^\prime Y & I^{\prime\prime} \, Y \! \! + \!
\frac{N_c}{3}
\end{array} \right)
\left( \begin{array}{cc||c} R_{B^\prime} & R_{\phi^\prime} & R_s \,
\gamma^\prime_s \\ I_{B^\prime} Y_{B^\prime} & I_{\phi^\prime}
Y_{\phi^\prime} & I_s Y_s \end{array} \right)
\nonumber \\ & & \times \sum_{K, \tilde{K} , \tilde{K}^\prime}
[K] ([\tilde{K}][\tilde{K}^\prime])^{1/2}
\left\{ \begin{array}{ccc}
L   & I                & \tilde{K} \\
S   & S_B              & S_\phi    \\
J_s & I^{\prime\prime} & K \end{array} \right\}
\! \left\{ \begin{array}{ccc}
L^\prime & I^\prime         & \tilde{K}^\prime \\
S^\prime & S_{B^\prime}     & S_{\phi^\prime} \\
J_s      & I^{\prime\prime} & K \end{array} \right\}
\tau^{\left\{ I I^\prime Y \right\}}_{K \tilde{K} \tilde{K}^\prime \!
L L^\prime} \ .
\label{Mmaster}
\end{eqnarray}
$S$ and $S^\prime$ indicate the total hadron spin angular momentum
({\it i.e.}, not including orbital angular momentum). The quantities
in braces are ordinary SU(2) $9j$ symbols.  The key quantum number
describing the dynamics of the reduced scattering amplitudes $\tau$ is
$K$, which in chiral soliton models represents the grand spin ${\bf K}
\! = \! {\bf I} \! + \! {\bf J}$.

In light of the results Eqs.~(\ref{dimsize}) and (\ref{CGCmag}),
Eq.~(\ref{Mmaster}) may be simplified considerably.  In particular,
Eq.~(\ref{CGCmag}) requires that the leading [$O(N_c^0)$] SU(3) CGC in
Eq.~(\ref{Mmaster}) have $Y \! = \! Y_\phi \! = \! Y_{\phi^\prime} \!
= \! r$, where $Y_{s, \rm max} \! = \! \frac{N_c}{3} \! + \! r$.  We
immediately see that the leading-order processes in $1/N_c$ require
$Y_B \! = \! Y_{B^\prime}$, {\it i.e.}, no strangeness change in the
scattered baryon, a fact that was used in Ref.~\cite{CLpenta}.  Also,
Eq.~(\ref{dimsize}) eliminates the SU(3) multiplet dimensions:
\begin{equation}
([R_B][R_{B^\prime}])^{1/2}/[R_s] \to
([S_B][S_{B^\prime}])^{1/2}/[I_{s,\rm{top}}] \ .
\end{equation}
Moreover, the product degeneracy factors $\gamma_{s,s^\prime}$ cannot
be discerned in any physical process and therefore must also be summed
over coherently in the full physical amplitude.

Specializing now to the case of nonstrange scattered baryons
($I_{B,B^\prime} \! = \! S_{B,B^\prime}$, $Y_{B,B^\prime} \! = \!
\frac{N_c}{3}$), one first notes that only intermediate states with
$Y_s \!  = \! Y_{s, \rm max}$ appear at $O(N_c^0)$.  In order to
recover the 2-flavor result, one must also note that implicit in
Eq.~(\ref{Mmaster}) is a factor $\delta_{R_s R^\prime_s}$, and that
the $R_s$ factor in the last two SU(3) CGC actually start as
$R^\prime_s$.  Then one sums over the intermediate SU(3) irreps $R_s$
and $R^\prime_s$.  Employing the well-known SU(3) CGC completeness
relation~\cite{deSwart}
\begin{equation}
\sum_{R \gamma Y}
\left( \begin{array}{cc||c} R_1 & R_2 & R \gamma \\ I_1
Y_1 & I_2 Y_2 & I Y \end{array} \right)
\left( \begin{array}{cc||c} R_1 & R_2 & R \gamma \\ I^\prime_1
Y^\prime_1 & I^\prime_2 Y^\prime_2 & I Y \end{array} \right)
= \delta_{I_1 I^\prime_1} \delta_{I_2 I^\prime_2} \delta_{Y_1
Y^\prime_1} \delta_{Y_2 Y^\prime_2} \, ,
\end{equation}
in the current case removes all SU(3) CGC and imposes $\delta_{I
I_\phi} \delta_{I^\prime I_{\phi^\prime}} \delta_{I^{\prime\prime}
I_s}$.  Noting that $I_\phi \! + \! Y_\phi/2$ and $I_{\phi^\prime} \!
+ \! Y_{\phi^\prime}/2$ are integers for mesons, one is left with the
2-flavor result~\cite{CLcompat,MMSU2},
\begin{eqnarray}
S_{L L^\prime S S^\prime I_s J_s} & = & \sum_{K, \tilde{K},
\tilde{K}^\prime} [K]
([S_B][S_{B^\prime}][S][S^\prime][\tilde{K}][\tilde{K}^\prime])^{1/2}
\nonumber \\ & &
\times \left\{ \begin{array}{ccc}
L & I_\phi & \tilde{K} \\
S & S_B & S_\phi \\
J_s & I_s & K \end{array} \right\} \left\{
\begin{array}{ccc}
L^\prime & I_{\phi^\prime} & \tilde{K}^\prime \\
S^\prime & S_{B^\prime} & S_{\phi^\prime} \\
J_s & I_s & K
\end{array}
\right\}
\tau_{K \tilde{K} \tilde{K}^\prime L L^\prime} \ ,
\label{2flavor}
\end{eqnarray}
where $\tau_{K \tilde{K} \tilde{K}^\prime L L^\prime} \! \equiv \!
(-1)^{I_B - I_{B^\prime} + I_\phi - I_{\phi^\prime}} \tau^{\{ I_\phi
I_{\phi^\prime} Y_\phi \}}_{K \tilde{K} \tilde{K}^\prime L L^\prime}$.

The phenomenologically most useful special case of Eq.~(\ref{Mmaster})
occurs for mesons in the $0^-$ octet, $S_\phi \! = \! S_{\phi^\prime}
\! = \! 0$.  Then the $9j$ symbols collapse to $6j$ symbols:
\begin{equation}
\left\{ \begin{array}{ccc}
L & I & \tilde{K} \\ S & S_B & 0 \\ J_s & I^{\prime\prime} & K
\end{array} \right\} = (-1)^{L + I^{\prime\prime} + K + S_B}
([S][\tilde{K}])^{-1/2} \delta_{S S_B}
\delta_{\tilde{K} K}
\sixj{K}{I^{\prime\prime}}{J_s}{S_B}{L}{I} \ ,
\end{equation}
and similarly for the other $9j$ symbol.  Then Eq.~(\ref{Mmaster})
simplifies by losing the sums on $\tilde{K}$ and $\tilde{K}^\prime$
and the $([S][S^\prime][\tilde{K}][\tilde{K}^\prime])^{1/2}$ factors,
as well as the phase $(-1)^{S_B - S_{B^\prime}}$ (These phases in
Eqs.~(2) and (6) of Ref.~\cite{CLSU3penta} are incorrect).  For
reference, the spinless meson expression reads
\begin{eqnarray}
\lefteqn{S_{L L^\prime S_B S_{B^\prime} J_s R_s \gamma_s
\gamma^\prime_s I_s Y_s}} \nonumber \\
& = &
(-1)^{L-L^\prime}
([R_B][R_B^\prime])^{1/2} / [R_s]
\sum_{\stackrel{\scriptstyle I, I^\prime \! , \, Y \in
8,}{I^{\prime\prime} \in R_s}}
(-1)^{I + I^\prime + Y} [I^{\prime\prime}]
\nonumber \\ & & \times
\left( \begin{array}{cc||c} R_B & 8 & R_s \, \gamma_s \\ S_B
\frac{N_c}{3} & I Y & I^{\prime\prime} \, Y \! \! + \! \frac{N_c}{3}
\end{array} \right)
\left( \begin{array}{cc||c} R_B & 8 & R_s \, \gamma_s \\ I_B
Y_B & I_\phi Y_\phi & I_s Y_s \end{array} \right) \nonumber \\
& & \times
\left( \begin{array}{cc||c} R_{B^\prime} & 8 & R_s \,
\gamma^\prime_s \\ S_{B^\prime}
\frac{N_c}{3} & I^\prime Y & I^{\prime\prime} \, Y \! \! + \!
\frac{N_c}{3} \end{array} \right)
\left( \begin{array}{cc||c} R_{B^\prime} & 8 & R_s \,
\gamma^\prime_s \\ I_{B^\prime} Y_{B^\prime} & I_{\phi^\prime}
Y_{\phi^\prime} & I_s Y_s \end{array} \right)
\nonumber \\ & & \times \sum_{K} [K]
\left\{ \begin{array}{ccc}
K   & I^{\prime\prime} & J_s \\
S_B & L                & I \end{array} \right\}
\! \left\{ \begin{array}{ccc}
K            & I^{\prime\prime} & J_s \\
S_{B^\prime} & L^\prime         & I^\prime \end{array} \right\}
\tau^{\left\{ I I^\prime Y \right\}}_{K K K L L^\prime} \ .
\label{spinless}
\end{eqnarray}
Note that, although we restrict to spinless mesons in this special
case, they are not necessarily pseudo-Goldstone bosons.  Chiral
symmetry is not imposed in any way and would provide additional
constraints.

The reduction of Eq.~(\ref{spinless}) to its nonstrange equivalent
works in precisely the same way as the reduction of
Eq.~(\ref{Mmaster}) to Eq.~(\ref{2flavor}).  One
finds~\cite{CL1st,MP}
\begin{eqnarray}
S_{LL^\prime S_B S_{B^\prime} I_s J_s}^{\phi \phi^\prime} & = &
(-1)^{S_{B^\prime} - S_B}
([S_B][S_{B^\prime}])^{1/2} \nonumber \\ & &
\sum_K [K] \left\{ \begin{array}{ccc} K &
I_s & J_s \\ S_{B^\prime} & L^\prime & I_{\phi^\prime} \end{array}
\right\} \left\{
\begin{array}{ccc} K & I_s & J_s \\ S_B & L & I_\phi \end{array}
\right\}
s_{KL^\prime L}^{\phi \phi^\prime} , \label{MPeqn1}
\end{eqnarray}
where $s^{\phi \phi^\prime}_{K L^\prime L} \! \equiv \! (-1)^{L -
L^\prime} \tau_{KKKLL^\prime} = (-1)^{L - L^\prime + I_B -
I_{B^\prime} + I_\phi - I_{\phi^\prime}} \tau^{\{I_\phi
I_{\phi^\prime} Y_\phi \}}_{KKKLL^\prime}$; for example,
$s^\pi_{KL^\prime L}$ in Ref.~\cite{CL1st} means $\phi \!  = \!
\phi^\prime \! = \pi$.

The notation for the full amplitudes is admittedly cumbersome due to
the numerous indices required for their unambiguous characterization.
The standard notation in the literature uses $L_{2I_s \, 2J_s}$ for
resonances with half-integer isospin, $L_{I_s \, 2J_s}$ otherwise.  Of
course, in reality almost all experiments scatter $0^-$ initial mesons
($\pi$'s and $K$'s) on nucleons ($J^P \! = \! {\frac 1 2}^+$), which
combined with parity conservation forces $L \! = \! L^\prime$ to be an
even integer.  In the following we are more interested in uncovering
the full pole structure than in presenting only expressions for $N_c
\! = \! 3$ physical amplitudes; we allow amplitudes with, for example,
an $\eta \Sigma^*$ initial state.  Nevertheless, as discussed below we
restrict to $B \! = \! B^\prime$, $\phi \! = \! \phi^\prime$, $L \! =
\! L^\prime$.  A sufficient notation for the amplitudes we present,
compatible with that used in
Refs.~\cite{CL1st,CLcompat,CLconfig,CL1N,CLphoto,CLpenta}, is
therefore $L^{\phi B}_{(2)I_s \, 2J_s}$.  The SU(3) labels $R_s
\gamma_s$ can also be made explicit if one is discussing a particular
resonant channel, but of course in real data these channels are summed
coherently.

\section{The Five Multiplets} \label{5mult}

In Ref.~\cite{CLSU3penta} we explained how Eq.~(\ref{spinless}) can be
used to uncover degenerate SU(3) multiplets of resonances.  In short,
one begins with a single established resonance with given $I_s, J_s$
quantum numbers, finds which reduced amplitude
$\tau_{KKKLL^\prime}^{\{II^\prime Y\}}$ contains the pole, notes that
the quantum numbers $LL^\prime$ and $II^\prime Y$ refer only to the
details of the formation of the resonance and not the resonance
itself, and concludes that resonance poles are labeled solely by $K$
[in the SU(3) limit].

We now show that the collection of resonance states with the quantum
numbers of the SU(6)$\times$O(3) (``{\bf 70}''$\!$,$\, 1^-$) multiplet
(the parity entering through allowed values of $L,L^\prime$) is
accommodated by 5 poles, one in each of 5 reduced amplitudes with $K
\! = \! 0, \frac 1 2, 1, \frac 3 2,$ and 2.

An exhaustive demonstration of this point would require the tabulation
of a huge set of amplitudes, including scattering with not only the
``{\bf 8}'' and ``{\bf 10}'', but also the stable baryons in the
``{\bf 56}'' with $J^P \!  = \!  {\frac 5 2}^+, \ldots ,
{\frac{N_c}{2}}^+$, as well as with members of the ``{\bf 8}'' and
``{\bf 10}'' carrying quantum numbers not appearing for $N_c \! = \!
3$ (such as an isospin-$\frac 3 2$ $\Xi$), or between states with $B
\! \neq B^\prime$, $\phi \! \neq \! \phi^\prime$, or $L \! \neq \!
L^\prime$ (Note however, that parity conservation in ``{\bf 8}''
${\frac 1 2}^+ \! \to \! {\frac 1 2}^+$ scattering with $0^-$ mesons
dictates $L \! = \! L^\prime$).  While many of these processes are
physically interesting ({\it e.g.}, $\pi N \to \eta N$), for our
purposes it is equally convincing to demonstrate the pole structure by
restricting the tabulation to a much smaller set: All quantum numbers
are chosen diagonal ($B \! = B^\prime$, $\phi \! = \! \phi^\prime$, $L
\!  = \!  L^\prime$), and only ``{\bf 8}''$\to$``{\bf 8}'' transitions
allowed for $N_c \! = \! 3$ are exhibited, except in the few
circumstances where ``{\bf 8}'' scattering does not access all the
poles, in which case ``{\bf 10}''$\to$``{\bf 10}'' scattering is also
exhibited.

The results of this analysis appear in Tables~\ref{t1}--\ref{t4}.
Note that the sign for the $\eta \, \Xi$(``{\bf 8}'')$\to \Xi$(``{\bf
8}$_2$'') CGC in Ref.~\cite{CLSU3} (relevant to Table~\ref{t2}) is
incorrect; it should begin with $+(3N_c \! - \! 19)$.  Also, the
symbol $\Omega^\prime$ (Table~\ref{t4}) indicates the $I \! = \! 1$
partner to the $\Omega$ for ``{\bf 10}'' with $N_c \! > \! 3$,
introduced to show that all expected resonance poles indeed occur.

In light of the fact that Eq.~(\ref{MPeqn1}) is a special case of the
full result Eq.~(\ref{Mmaster}), the results appearing in Table~I of
Ref.~\cite{CL1st} still hold, demonstrating that all nonstrange
resonances in the ``{\bf 70}'' reduce to 3 poles with $K \! = \!  0,
1,$ and 2.  Then, since Eq.~(\ref{Mmaster}) is an SU(3)-symmetric
expression, the same pole that produces a given nonstrange resonance
must also produce all its SU(3)-multiplet partners.  For example, the
$N_{1/2}$ state corresponding to the $K \! = \! 0$ pole is but the
nonstrange member of an ``{\bf 8}'' corresponding to the same pole.
This point is also apparent in the tables.

\begin{table}
\caption{Partial-wave amplitudes containing resonances with quantum
numbers corresponding to states in the large $N_c$ quark-picture
SU(6)$\times$O(3) (``{\bf 70}''$\!$,$\, 1^-$).  Expansions are given
in terms of $K$-amplitudes, according to Eq.~(\ref{spinless}).
\label{t1}}
\begin{tabular}{lcccccl}
State \mbox{  } && ``{\bf 70}'' Pole Masses \mbox{   } &&
\multicolumn{3}{l}{Partial Wave, $K$-Amplitudes} \\
\hline\hline
$\Lambda_{1/2} \mbox{(``{\bf 8}'')}$ && $m_0$, $m_1$
   && $S^{\pi \Sigma}_{01}$   &=& $\tau^{\{110\}}_{11100}$ \\
&& && $S^{\eta \Lambda}_{01}$ &=& $\tau^{\{000\}}_{00000}$ \\
\hline
$\Lambda_{3/2} \mbox{(``{\bf 8}'')}$ && $m_1$, $m_2$
   && $D^{\pi \Sigma}_{03}$   &=& $\frac 1 2
( \tau^{\{110\}}_{11122} + \tau^{\{110\}}_{22222} )$ \\
&& && $D^{\eta \Lambda}_{03}$ &=& $\tau^{\{000\}}_{22222}$ \\
\hline
$\Lambda_{5/2} \mbox{(``{\bf 8}'')}$ && $m_2$
   && $D^{\pi \Sigma}_{05}$   &=& $\frac 1 9
( 2\tau^{\{110\}}_{22222} + 7\tau^{\{110\}}_{33322} )$ \\
&& && $D^{\eta \Lambda}_{05}$ &=& $\tau^{\{000\}}_{22222}$ \\
\hline
$\Lambda_{1/2} \mbox{(``{\bf 1}'')}$ && $m_{\frac 1 2}$
   && $S^{\overline{K} N}_{01}$ &=& $\tau^{\{ \frac 1 2
\frac 1 2 \, -1\}}_{\frac 1 2 \frac 1 2 \frac 1 2 0 0}$ \\
\hline
$\Lambda_{3/2} \mbox{(``{\bf 1}'')}$ && $m_{\frac 3 2}$
   && $D^{\overline{K} N}_{03}$            &=& $\tau^{\{ \frac 1 2
\frac 1 2 \, -1 \}}_{\frac 3 2 \frac 3 2 \frac 3 2 2 2}$ \\
\hline
$\Sigma_{1/2} \mbox{(``{\bf 8}'')}$ && $m_0$, $m_1$
   && $S^{\pi \Lambda}_{11}$ &=&
$\frac 1 3 \tau^{\{ 110 \}}_{11100}$ \\
&& && $S^{\pi \Sigma}_{11}$  &=&
$\frac 2 3 \tau^{\{ 110 \}}_{11100}$ \\
&& && $S^{\eta \Sigma}_{11}$ &=&
$\tau^{\{ 000 \}}_{00000}$ \\
\hline
$\Sigma_{3/2} \mbox{(``{\bf 8}'')}$  && $m_1$, $m_2$
   && $D^{\pi \Lambda}_{13}$ &=&
$\frac 1 6 ( \tau^{\{ 110 \}}_{11122} + \tau^{\{ 110 \}}_{22222} )$ \\
&& && $D^{\pi \Sigma}_{13}$  &=&
$\frac 1 3 ( \tau^{\{ 110 \}}_{11122} + \tau^{\{ 110 \}}_{22222} )$ \\
&& && $D^{\eta \Sigma}_{13}$ &=& $\tau^{\{ 000 \}}_{22222}$ \\
\hline
$\Sigma_{5/2} \mbox{(``{\bf 8}'')}$  && $m_2$
   && $D^{\pi \Lambda}_{15}$ &=& $\frac{1}{27}
( 2 \tau^{\{ 110 \}}_{22222} + 7 \tau^{\{ 110 \}}_{33322} )$ \\
&& && $D^{\pi \Sigma}_{15}$  &=& $\frac{2}{27}
( 2 \tau^{\{ 110 \}}_{22222} + 7 \tau^{\{ 110 \}}_{33322} )$ \\
&& && $D^{\eta \Sigma}_{15}$ &=& $\tau^{\{ 000 \}}_{22222}$ \\
\hline
$\Sigma_{1/2} \mbox{(``{\bf S}'')}$ && $m_{\frac 1 2}$, $m_{\frac 3 2}$
   && $S^{\overline{K} N}_{11}$ &=&
$\tau^{\{ \frac 1 2 \frac 1 2 \, -1 \}}
_{\frac 1 2 \frac 1 2 \frac 1 2 00}$ \\
&& && $D^{\overline{K} \Delta}_{11}$  &=&
$\tau^{\{ \frac 1 2 \frac 1 2 \, -1 \}}
_{\frac 3 2 \frac 3 2 \frac 3 2 22}$ \\
\hline
$\Sigma_{3/2} \mbox{(``{\bf S}'')}$ && $m_{\frac 1 2}$, $m_{\frac 3 2}$
   && $D^{\overline{K} N}_{13}$ &=&
$\frac 1 5 ( \tau^{\{ \frac 1 2 \frac 1 2 \, -1 \}}
_{\frac 3 2 \frac 3 2 \frac 3 2 22} +
4 \tau^{\{ \frac 1 2 \frac 1 2 \, -1 \}}
_{\frac 5 2 \frac 5 2 \frac 5 2 22} )$ \\
&& && $S^{\overline{K} \Delta}_{13}$  &=&
$\tau^{\{ \frac 1 2 \frac 1 2 \, -1 \}}
_{\frac 1 2 \frac 1 2 \frac 1 2 00}$ \\
&& && $D^{\overline{K} \Delta}_{13}$  &=&
$\frac 1 5 ( 4 \tau^{\{ \frac 1 2 \frac 1 2 \, -1 \}}
_{\frac 3 2 \frac 3 2 \frac 3 2 22} +
\tau^{\{ \frac 1 2 \frac 1 2 \, -1 \}}
_{\frac 5 2 \frac 5 2 \frac 5 2 22} )$ \\
\hline
$\Sigma_{5/2} \mbox{(``{\bf S}'')}$ && $m_{\frac 3 2}$
   && $D^{\overline{K} N}_{15}$ &=&
$\frac{1}{15} ( 8 \tau^{\{ \frac 1 2 \frac 1 2 \, -1 \}}
_{\frac 3 2 \frac 3 2 \frac 3 2 22} +
7 \tau^{\{ \frac 1 2 \frac 1 2 \, -1 \}}
_{\frac 5 2 \frac 5 2 \frac 5 2 22} )$ \\
&& && $D^{\overline{K} \Delta}_{15}$  &=&
$\frac{1}{15} ( 7 \tau^{\{ \frac 1 2 \frac 1 2 \, -1 \}}
_{\frac 3 2 \frac 3 2 \frac 3 2 22} +
8 \tau^{\{ \frac 1 2 \frac 1 2 \, -1 \}}
_{\frac 5 2 \frac 5 2 \frac 5 2 22} )$ \\
&& && $G^{\overline{K} \Delta}_{15}$  &=&
$\tau^{\{ \frac 1 2 \frac 1 2 \, -1 \}}
_{\frac 7 2 \frac 7 2 \frac 7 2 44}$ \\
\hline
\end{tabular}
\end{table}

\begin{table}
\caption{First continuation of Table~\ref{t1}.\label{t2}}
\begin{tabular}{lcccccl}
State \mbox{  } && ``{\bf 70}'' Pole Masses \mbox{   } &&
\multicolumn{3}{l}{Partial Wave, $K$-Amplitudes} \\
\hline\hline
$\Sigma_{1/2} \mbox{(``{\bf 10}'')}$ && $m_1$, $m_2$
   && $S^{\pi \Lambda}_{11}$ &=&
$\frac 2 3 \tau^{\{ 110 \}}_{11100}$ \\
&& && $S^{\pi \Sigma}_{11}$  &=&
$\frac 1 3 \tau^{\{ 110 \}}_{11100}$ \\
&& && $D^{\eta \Sigma^*}_{11}$ &=&
$\tau^{\{ 000 \}}_{22222}$ \\
\hline
$\Sigma_{3/2} \mbox{(``{\bf 10}'')}$ && $m_0$, $m_1$, $m_2$
   && $D^{\pi \Lambda}_{13}$ &=& $\frac{1}{30}
( \tau^{\{ 110 \}}_{11122} + 5 \tau^{\{110\}}_{22222} + 14
\tau^{\{ 110 \}}_{33322} )$ \\
&& && $D^{\pi \Sigma}_{13}$  &=& $\frac{1}{60}
( \tau^{\{ 110 \}}_{11122} + 5 \tau^{\{110\}}_{22222} + 14
\tau^{\{ 110 \}}_{33322} )$ \\
&& && $S^{\eta \Sigma^*}_{13}$ &=& $\tau^{\{ 000 \}}_{00000}$ \\
&& && $D^{\eta \Sigma^*}_{13}$ &=& $\tau^{\{ 000 \}}_{22222}$ \\
\hline
$\Sigma_{5/2} \mbox{(``{\bf 10}'')}$ && $m_1$, $m_2$
   && $D^{\pi \Lambda}_{15}$ &=& $\frac{1}{135}
( 27 \tau^{\{ 110 \}}_{11122} + 35 \tau^{\{ 110 \}}_{22222} +
28 \tau^{\{ 110 \}}_{33322} )$ \\
&& && $D^{\pi \Sigma}_{15}$  &=& $\frac{1}{270}
( 27 \tau^{\{ 110 \}}_{11122} + 35 \tau^{\{ 110 \}}_{22222} +
28 \tau^{\{ 110 \}}_{33322} )$ \\
&& && $D^{\eta \Sigma^*}_{15}$ &=& $\tau^{\{ 000 \}}_{22222}$ \\
&& && $G^{\eta \Sigma^*}_{15}$ &=& $\tau^{\{ 000 \}}_{44444}$ \\
\hline
$\Sigma_{7/2} \mbox{(``{\bf 10}'')}$ && $m_2$
   && $G^{\pi \Lambda}_{17}$ &=& $\frac{1}{108}
( 7 \tau^{\{ 110 \}}_{33344} + 21 \tau^{\{ 110 \}}_{44444} +
44 \tau^{\{ 110 \}}_{55544} )$ \\
&& && $G^{\pi \Sigma}_{17}$ &=& $\frac{1}{216}
( 7 \tau^{\{ 110 \}}_{33344} + 21 \tau^{\{ 110 \}}_{44444} +
44 \tau^{\{ 110 \}}_{55544} )$ \\
&& && $D^{\eta \Sigma^*}_{17}$ &=& $\tau^{\{ 000 \}}_{22222}$ \\
&& && $G^{\eta \Sigma^*}_{17}$ &=& $\tau^{\{ 000 \}}_{44444}$ \\
\hline
$\Xi_{1/2} \mbox{(``{\bf 8}'')}$ && $m_0$, $m_1$
   && $S^{\pi \Xi}_{11}$  &=& $\frac 1 9 \tau^{\{ 110 \}}_{11100}$ \\
&& && $S^{\eta \Xi}_{11}$ &=& $\tau^{\{ 000 \}}_{00000}$ \\
\hline
$\Xi_{3/2} \mbox{(``{\bf 8}'')}$ && $m_1$, $m_2$
   && $D^{\pi \Xi}_{13}$ &=&
$\frac{1}{18} ( \tau^{\{ 110 \}}_{11122} + \tau^{\{ 110 \}}_{22222} )$ \\
&& && $D^{\eta \Xi}_{13}$  &=& $\tau^{\{ 000 \}}_{22222}$ \\
\hline
$\Xi_{5/2} \mbox{(``{\bf 8}'')}$  && $m_2$
   && $D^{\pi \Xi}_{15}$ &=&
$\frac{1}{81} ( 2 \tau^{\{ 110 \}}_{22222} +
7 \tau^{\{ 110 \}}_{33322} )$ \\
&& && $D^{\eta \Xi}_{15}$  &=& $\tau^{\{ 000 \}}_{22222}$ \\
\hline
$\Xi_{1/2} \mbox{(``{\bf 1}'')}$ && $m_{\frac 1 2}$
   && $S^{\overline{K} \Lambda}_{11}$  &=&
$\frac 1 4 \tau^{\{ \frac 1 2 \frac 1 2 \, -1 \}}
_{\frac 1 2 \frac 1 2 \frac 1 2 00}$ \\
&& && $S^{\overline{K} \Sigma}_{11}$  &=&
$\frac 3 4 \tau^{\{ \frac 1 2 \frac 1 2 \, -1 \}}
_{\frac 1 2 \frac 1 2 \frac 1 2 00}$ \\
\hline
$\Xi_{3/2} \mbox{(``{\bf 1}'')}$ && $m_{\frac 3 2}$
   && $D^{\overline{K} \Lambda}_{13}$ &=&
$\frac 1 4 \tau^{\{ \frac 1 2 \frac 1 2 \, -1 \}}
_{\frac 3 2 \frac 3 2 \frac 3 2 22}$ \\
&& && $D^{\overline{K} \Sigma}_{13}$  &=&
$\frac 3 4 \tau^{\{ \frac 1 2 \frac 1 2 \, -1 \}}
_{\frac 3 2 \frac 3 2 \frac 3 2 22}$ \\
\hline
\end{tabular}
\end{table}

\begin{table}
\caption{Second continuation of Table~\ref{t1}.\label{t3}}
\begin{tabular}{lcccccl}
State \mbox{ } && ``{\bf 70}'' Pole Masses \mbox{ } &&
\multicolumn{3}{l}{Partial Wave, $K$-Amplitudes} \\
\hline\hline
$\Xi_{1/2} \mbox{(``{\bf 10}'')}$ && $m_1$, $m_2$
   && $S^{\pi \Xi}_{11}$ &=&
$\frac 8 9 \tau^{\{ 110 \}}_{11100}$ \\
&& && $D^{\eta \Xi^*}_{11}$  &=&
$\tau^{\{ 000 \}}_{22222}$ \\
\hline
$\Xi_{3/2} \mbox{(``{\bf 10}'')}$ && $m_0$, $m_1$, $m_2$
   && $D^{\pi \Xi}_{13}$ &=& $\frac{2}{45}
( \tau^{\{ 110 \}}_{11122} + 5 \tau^{\{110\}}_{22222} + 14
\tau^{\{ 110 \}}_{33322} )$ \\
&& && $S^{\eta \Xi^*}_{13}$ &=& $\tau^{\{ 000 \}}_{00000}$ \\
&& && $D^{\eta \Xi^*}_{13}$ &=& $\tau^{\{ 000 \}}_{22222}$ \\
\hline
$\Xi_{5/2} \mbox{(``{\bf 10}'')}$ && $m_1$, $m_2$
   && $D^{\pi \Xi}_{15}$ &=& $\frac{4}{405}
( 27 \tau^{\{ 110 \}}_{11122} + 35 \tau^{\{ 110 \}}_{22222} +
28 \tau^{\{ 110 \}}_{33322} )$ \\
&& && $D^{\eta \Xi^*}_{15}$ &=& $\tau^{\{ 000 \}}_{22222}$ \\
&& && $G^{\eta \Xi^*}_{15}$ &=& $\tau^{\{ 000 \}}_{44444}$ \\
\hline
$\Xi_{7/2} \mbox{(``{\bf 10}'')}$ && $m_2$
   && $G^{\pi \Xi}_{17}$ &=& $\frac{1}{81}
( 7 \tau^{\{ 110 \}}_{33344} + 21 \tau^{\{ 110 \}}_{44444} +
44 \tau^{\{ 110 \}}_{55544} )$ \\
&& && $D^{\eta \Xi^*}_{17}$ &=& $\tau^{\{ 000 \}}_{22222}$ \\
&& && $G^{\eta \Xi^*}_{17}$ &=& $\tau^{\{ 000 \}}_{44444}$ \\
\hline
$\Xi_{1/2} \mbox{(``{\bf S}'')}$ && $m_{\frac 1 2}$, $m_{\frac 3 2}$
   && $S^{\overline{K} \Lambda}_{11}$  &=&
$\frac 3 4 \tau^{\{ \frac 1 2 \frac 1 2 \, -1 \}}
_{\frac 1 2 \frac 1 2 \frac 1 2 00}$ \\
&& && $S^{\overline{K} \Sigma}_{11}$ &=&
$\frac 1 4 \tau^{\{ \frac 1 2 \frac 1 2 \, -1 \}}
_{\frac 1 2 \frac 1 2 \frac 1 2 00}$ \\
&& && $D^{\overline{K} \Sigma^*}_{11}$  &=&
$\tau^{\{ \frac 1 2 \frac 1 2 \, -1 \}}
_{\frac 3 2 \frac 3 2 \frac 3 2 22}$ \\
\hline
$\Xi_{3/2} \mbox{(``{\bf S}'')}$ && $m_{\frac 1 2}$, $m_{\frac 3 2}$
   && $D^{\overline{K} \Lambda}_{13}$ &=&
$\frac{3}{20} ( \tau^{\{ \frac 1 2 \frac 1 2 \, -1 \}}
_{\frac 3 2 \frac 3 2 \frac 3 2 22} +
4 \tau^{\{ \frac 1 2 \frac 1 2 \, -1 \}}
_{\frac 5 2 \frac 5 2 \frac 5 2 22} )$ \\
&& && $D^{\overline{K} \Sigma}_{13}$  &=&
$\frac{1}{20} ( \tau^{\{ \frac 1 2 \frac 1 2 \, -1 \}}
_{\frac 3 2 \frac 3 2 \frac 3 2 22} +
4 \tau^{\{ \frac 1 2 \frac 1 2 \, -1 \}}
_{\frac 5 2 \frac 5 2 \frac 5 2 22} )$ \\
&& && $S^{\overline{K} \Sigma^*}_{13}$  &=&
$\tau^{\{ \frac 1 2 \frac 1 2 \, -1 \}}
_{\frac 1 2 \frac 1 2 \frac 1 2 00}$ \\
&& && $D^{\overline{K} \Sigma^*}_{13}$  &=&
$\frac 1 5 ( 4 \tau^{\{ \frac 1 2 \frac 1 2 \, -1 \}}
_{\frac 3 2 \frac 3 2 \frac 3 2 22} +
\tau^{\{ \frac 1 2 \frac 1 2 \, -1 \}}
_{\frac 5 2 \frac 5 2 \frac 5 2 22} )$ \\
\hline
$\Xi_{5/2} \mbox{(``{\bf S}'')}$  && $m_{\frac 3 2}$
   && $D^{\overline{K} \Lambda}_{15}$ &=&
$\frac{1}{20} ( 8 \tau^{\{ \frac 1 2 \frac 1 2 \, -1 \}}
_{\frac 3 2 \frac 3 2 \frac 3 2 22} +
7 \tau^{\{ \frac 1 2 \frac 1 2 \, -1 \}}
_{\frac 5 2 \frac 5 2 \frac 5 2 22} )$ \\
&& && $D^{\overline{K} \Sigma}_{15}$  &=&
$\frac{1}{60} ( 8 \tau^{\{ \frac 1 2 \frac 1 2 \, -1 \}}
_{\frac 3 2 \frac 3 2 \frac 3 2 22} +
7 \tau^{\{ \frac 1 2 \frac 1 2 \, -1 \}}
_{\frac 5 2 \frac 5 2 \frac 5 2 22} )$ \\
&& && $D^{\overline{K} \Sigma^*}_{15}$  &=&
$\frac{1}{15} ( 7 \tau^{\{ \frac 1 2 \frac 1 2 \, -1 \}}
_{\frac 3 2 \frac 3 2 \frac 3 2 22} +
8 \tau^{\{ \frac 1 2 \frac 1 2 \, -1 \}}
_{\frac 5 2 \frac 5 2 \frac 5 2 22} )$ \\
&& && $G^{\overline{K} \Sigma^*}_{15}$  &=&
$\tau^{\{ \frac 1 2 \frac 1 2 \, -1 \}}
_{\frac 7 2 \frac 7 2 \frac 7 2 44}$ \\
\hline
\end{tabular}
\end{table}

\begin{table}
\caption{Third continuation of Table~\ref{t1}. $\Omega^\prime$ is the
$I \! = \! 1$ partner of the $\Omega$ in ``{\bf 10}'' for $N_c \! > \!
3$.\label{t4}}
\begin{tabular}{lcccccl}
State \mbox{  } && ``{\bf 70}'' Pole Masses \mbox{   } &&
\multicolumn{3}{l}{Partial Wave, $K$-Amplitudes} \\
\hline\hline
$\Omega_{1/2} \mbox{(``{\bf 10}'')}$ && $m_1$, $m_2$
   && $S^{\pi \Omega^\prime}_{01}$ &=&
$\tau^{\{ 110 \}}_{11100}$ \\
&& && $D^{\eta \Omega}_{01}$  &=&
$\tau^{\{ 000 \}}_{22222}$ \\
\hline
$\Omega_{3/2} \mbox{(``{\bf 10}'')}$ && $m_0$, $m_1$, $m_2$
   && $D^{\pi \Omega^\prime}_{03}$ &=& $\frac{1}{20}
( \tau^{\{ 110 \}}_{11122} + 5 \tau^{\{110\}}_{22222} + 14
\tau^{\{ 110 \}}_{33322} )$ \\
&& && $S^{\eta \Omega}_{03}$ &=& $\tau^{\{ 000 \}}_{00000}$ \\
&& && $D^{\eta \Omega}_{03}$ &=& $\tau^{\{ 000 \}}_{22222}$ \\
\hline
$\Omega_{5/2} \mbox{(``{\bf 10}'')}$ && $m_1$, $m_2$
   && $D^{\pi \Omega^\prime}_{05}$ &=& $\frac{1}{90}
( 27 \tau^{\{ 110 \}}_{11122} + 35 \tau^{\{ 110 \}}_{22222} +
28 \tau^{\{ 110 \}}_{33322} )$ \\
&& && $D^{\eta \Omega}_{05}$ &=& $\tau^{\{ 000 \}}_{22222}$ \\
&& && $G^{\eta \Omega}_{05}$ &=& $\tau^{\{ 000 \}}_{44444}$ \\
\hline
$\Omega_{7/2} \mbox{(``{\bf 10}'')}$ && $m_2$
   && $G^{\pi \Omega^\prime}_{07}$ &=& $\frac{1}{72}
( 7 \tau^{\{ 110 \}}_{33344} + 21 \tau^{\{ 110 \}}_{44444} +
44 \tau^{\{ 110 \}}_{55544} )$ \\
&& && $D^{\eta \Omega}_{07}$ &=& $\tau^{\{ 000 \}}_{22222}$ \\
&& && $G^{\eta \Omega}_{07}$ &=& $\tau^{\{ 000 \}}_{44444}$ \\
\hline
$\Omega_{1/2} \mbox{(``{\bf S}'')}$ && $m_{\frac 1 2}$, $m_{\frac 3 2}$
   && $S^{\overline{K} \Xi}_{01}$  &=&
$\tau^{\{ \frac 1 2 \frac 1 2 \, -1 \}}
_{\frac 1 2 \frac 1 2 \frac 1 2 00}$ \\
&& && $D^{\overline{K} \Xi^*}_{01}$  &=&
$\tau^{\{ \frac 1 2 \frac 1 2 \, -1 \}}
_{\frac 3 2 \frac 3 2 \frac 3 2 22}$ \\
\hline
$\Omega_{3/2} \mbox{(``{\bf S}'')}$ && $m_{\frac 1 2}$, $m_{\frac 3 2}$
   && $D^{\overline{K} \Xi}_{03}$ &=&
$\frac{1}{5} ( \tau^{\{ \frac 1 2 \frac 1 2 \, -1 \}}
_{\frac 3 2 \frac 3 2 \frac 3 2 22} +
4 \tau^{\{ \frac 1 2 \frac 1 2 \, -1 \}}
_{\frac 5 2 \frac 5 2 \frac 5 2 22} )$ \\
&& && $S^{\overline{K} \Xi^*}_{03}$  &=&
$\tau^{\{ \frac 1 2 \frac 1 2 \, -1 \}}
_{\frac 1 2 \frac 1 2 \frac 1 2 00}$ \\
&& && $D^{\overline{K} \Xi^*}_{03}$  &=&
$\frac 1 5 ( 4 \tau^{\{ \frac 1 2 \frac 1 2 \, -1 \}}
_{\frac 3 2 \frac 3 2 \frac 3 2 22} +
\tau^{\{ \frac 1 2 \frac 1 2 \, -1 \}}
_{\frac 5 2 \frac 5 2 \frac 5 2 22} )$ \\
\hline
$\Omega_{5/2} \mbox{(``{\bf S}'')}$  && $m_{\frac 3 2}$
   && $D^{\overline{K} \Xi}_{05}$ &=&
$\frac{1}{15} ( 8 \tau^{\{ \frac 1 2 \frac 1 2 \, -1 \}}
_{\frac 3 2 \frac 3 2 \frac 3 2 22} +
7 \tau^{\{ \frac 1 2 \frac 1 2 \, -1 \}}
_{\frac 5 2 \frac 5 2 \frac 5 2 22} )$ \\
&& && $D^{\overline{K} \Xi^*}_{05}$  &=&
$\frac{1}{15} ( 7 \tau^{\{ \frac 1 2 \frac 1 2 \, -1 \}}
_{\frac 3 2 \frac 3 2 \frac 3 2 22} +
8 \tau^{\{ \frac 1 2 \frac 1 2 \, -1 \}}
_{\frac 5 2 \frac 5 2 \frac 5 2 22} )$ \\
&& && $G^{\overline{K} \Xi^*}_{05}$  &=&
$\tau^{\{ \frac 1 2 \frac 1 2 \, -1 \}}
_{\frac 7 2 \frac 7 2 \frac 7 2 44}$ \\
\hline
\end{tabular}
\end{table}

One concludes from studying Tables~\ref{t1}--\ref{t4} that the 20
SU(3) multiplets of the (``{\bf 70}''$\!$,$\, 1^-$) listed in
Eq.~(\ref{70mult}) actually collect into 5 irreps labeled by $K$:
\begin{eqnarray}
K = 0 & : &
\left( \mbox{``{\bf 8}''}, \frac 1 2 \right) \oplus
\left( \mbox{``{\bf 10}''}, \frac 3 2 \right) \ , \nonumber \\
K = \frac 1 2 & : &
\left( \mbox{``{\bf 1}''}, \frac 1 2 \right) \oplus
\left( \mbox{``{\bf S}''}, \frac 1 2 \right) \oplus
\left( \mbox{``{\bf S}''}, \frac 3 2 \right) \ , \nonumber \\
K = 1 & : &
\left( \mbox{``{\bf 8}''}, \frac 1 2 \right) \oplus
\left( \mbox{``{\bf 8}''}, \frac 3 2 \right) \oplus
\left( \mbox{``{\bf 10}''}, \frac 1 2 \right) \oplus
\left( \mbox{``{\bf 10}''}, \frac 3 2 \right) \oplus
\left( \mbox{``{\bf 10}''}, \frac 5 2 \right) \ , \nonumber \\
K = \frac 3 2 & : &
\left( \mbox{``{\bf 1}''}, \frac 3 2 \right) \oplus
\left( \mbox{``{\bf S}''}, \frac 1 2 \right) \oplus
\left( \mbox{``{\bf S}''}, \frac 3 2 \right) \oplus
\left( \mbox{``{\bf S}''}, \frac 5 2 \right) \ , \nonumber \\
K = 2 & : &
\left( \mbox{``{\bf 8}''}, \frac 3 2 \right) \oplus
\left( \mbox{``{\bf 8}''}, \frac 5 2 \right) \oplus
\left( \mbox{``{\bf 10}''}, \frac 1 2 \right) \oplus
\left( \mbox{``{\bf 10}''}, \frac 3 2 \right) \oplus
\left( \mbox{``{\bf 10}''}, \frac 5 2 \right) \oplus
\left( \mbox{``{\bf 10}''}, \frac 7 2 \right) \ . \nonumber \\ &&
\label{Kmult}
\end{eqnarray}

That the large $N_c$ quark model SU(6)$\times$O(3) (``{\bf
70}''$\!$,$\, 1^-$) multiplet actually contains 5 independent mass
eigenvalues split by $O(N_c^0)$ can also be seen by referring to the
Hamiltonian operator basis used in Refs.~\cite{nonstrange,strange}.
This analysis extends that performed in Ref.~\cite{CL1st} for the
nonstrange case, in which one finds 3 operators with linearly
independent matrix elements up to $O(N_c^0)$, and only 3 distinct mass
eigenvalues.  By direct construction, one finds a single operator,
$O_1 \! = \! \openone$, whose matrix elements on all baryons is
precisely $N_c$, and 4 operators with $O(N_c^0)$ matrix elements:
\begin{equation} \label{Nc0ops}
O_2 = \ell s, \ O_3 = \frac{1}{N_c} \ell^{(2)} g G_c, \ O_4 =
\ell s + \frac{4}{N_c + 1} \ell t G_c, \
O_5 = \frac{1}{N_c} \left( tT_c - \frac{1}{12} \openone \right) \ .
\end{equation}
Here, $\ell$ is the orbital excitation operator, while $\ell^{(2)}$ is
the $\Delta \ell \! = \! 2$ tensor operator $(\ell^i \ell^j \! - \frac
1 3 \delta^{ij} \ell^2)$.  Lowercase indicates operators acting upon
the excited quark, and uppercase (with subscript $c$) indicates
operators acting upon the core.  $S$, $T$, and $G$ denote operators
with spin, flavor, and both spin and flavor indices, respectively,
summed over all relevant quarks, and all spin and flavor indices
implied by the component operators in Eq.~(\ref{Nc0ops}) are summed in
the unique nontrivial manner and then suppressed ({\it e.g.}, $\ell t
G_c \! \equiv \! \ell^i t^a G_c^{ia}$).  The operators are equivalent
to those at $O(N_c^0)$ in Ref.~\cite{strange}, except for the addition
of $O_5$, which was omitted in that work~\cite{GSSdisc}.  $O_4$ and
$O_5$ appear in more complicated forms whose matrix elements vanish
for all states in multiplets with $Y_{\rm max} \! = \! \frac{N_c}{3}$
(which includes all nonstrange states in the ``{\bf 70}''); in
Ref.~\cite{nonstrange} they were termed ``demotable''.  For the
$Y_{\rm max} \! = \! \frac{N_c }{3} \! - \! 1$ multiplets (``{\bf 1}''
and ``{\bf S}''), the matrix elements of $O_5$ at $O(N_c^0)$ are found
to be $-\frac 1 4$.

Using this notation, one finds that each mass eigenstate in the ``{\bf
70}'' assumes one of only 5 distinct eigenvalues.  Those in the SU(3)
multiplets with $Y \! = \! \frac{N_c}{3}$ assume the values
\begin{eqnarray}
m_0 & \equiv & c_1 N_c - \left( c_2 + \frac{5}{24} c_3 \right) \ ,
\nonumber \\
m_1 & \equiv & c_1 N_c - \frac 1 2 \left( c_2 - \frac{5}{24} c_3
\right) \ , \nonumber \\
m_2 & \equiv & c_1 N_c + \frac 1 2 \left( c_2 -
\frac{1}{24} c_3 \right) \ ,
\end{eqnarray}
which are the same expressions as in Ref.~\cite{CL1st}.  For the
$Y_{\rm max} \! = \! \frac{N_c }{3} - \! 1$ multiplets, one
additionally finds only the eigenvalues
\begin{eqnarray}
m_{\frac 1 2} & \equiv & c_1 N_c - (c_2 + c_4) -\frac 1 4 c_5 \ ,
\nonumber \\
m_{\frac 3 2} & \equiv & c_1 N_c + \frac 1 2 (c_2 + c_4) - \frac 1 4
c_5 \ .
\end{eqnarray}
Again, one sees that the (``{\bf 70}''$\!$,$\, 1^-$) is actually a
reducible collection of 5 multiplets.  The mass eigenvalues, labeled
by $m_K$, are listed in Tables~\ref{t1}--\ref{t4}.  While the old
SU(6) symmetry does not hold at $O(N_c^0)$, the remaining level of
degeneracy remains remarkable; for example, the multiplets listed in
Eq.~(\ref{Kmult}) mean that 5 eigenvalues [in the SU(3) limit] span 71
distinct isomultiplets, 30 for $N_c \! = \! 3$.  And even when SU(3)
symmetry is arbitrarily broken ({\it i.e.}, reduced amplitudes $\tau$
with the same $K$ but different $Y$ are taken to be distinct),
isomultiplets with the same value of $Y$ and $K$ but in different
SU(3) irreps in Eq.~(\ref{Kmult}) remain degenerate.

\section{Phenomenological Results} \label{phenom}

In this section we combine our qualitative results with experimental
extractions of BR to determine phenomenologically the SU(3) and $K$
irreducible representations (irreps) of various excited baryons.  This
is useful for two reasons.  First, it gives insight into the nature of
these resonant states in a framework independent of the quark model.
Second, the extent to which the decays fall into patterns consistent
with large $N_c$ predictions provides a check on the applicability of
the large $N_c$ approach to excited states for the real world of $N_c
\! = \!  3$.

Before discussing individual states in detail a few comments are in
order.  As noted in the Introduction, the extraction of BR necessarily
involves some modeling.  In some cases the model dependence is small,
and robust extractions of BR are possible.  However, in many cases
either the model dependence is large or the experimental data is
insufficient, and the BR are not known well.  Often the ranges for BR
quoted by the Particle Data Group are quite broad~\cite{PDG}.  Indeed,
they are often so large that it is impossible to make even qualitative
assessments of the dominant mode of decay.  Accordingly, we focus our
attention on those cases where the BR are relatively well established.

Another issue that should kept in mind in this discussion is that the
analysis presented so far is based on exact SU(3) flavor symmetry.  Of
course, in the real world SU(3) flavor is broken.  The analysis is
useful provided that SU(3) flavor violations are relatively modest
(which they usually are).  Similarly, the analysis is based on large
$N_c$ and implicitly assumes that $1/N_c$ corrections are small.
However, in one obvious case both SU(3) violations and $1/N_c$
corrections can be expected to be greatly enhanced: resonant states
not far above thresholds.  In such regions the phase space is a very
sensitive function of the masses, and a relatively small mass change
can lead to dramatic shifts in the phase space.  This near-threshold
behavior is particularly critical in high $L$ partial waves, where the
partial decay rate scales as $p^{2 L + 1}$, $p$ being the 3-momentum
of either outgoing particle in the center-of-momentum frame.
Accordingly, we focus on large $N_c$ predictions for the coupling in
various decays rather than on the partial widths or BR, since the
couplings are far less sensitive to threshold effects.

The most striking result of this work has already been described in
Sec.~\ref{CGCthm} and particularly by the constraint
Eq.~(\ref{CGCmag}): In the large $N_c$ limit, baryon resonances couple
only to mesons with a hypercharge equal to the amount by which the
hypercharge of the top row of its SU(3) multiplet exceeds that of the
stable baryons ($Y_{\rm max} \! = \frac{N_c}{3}$).  This result
provides a means by which the singlet and octet $\Lambda$ may be
distinguished: The former prefers $\overline{K} N$ to $\pi \Sigma$
decays with a coupling $O(N_c)$ larger than predicted by phase space
alone, and vice versa.

One sees this effect clearly in some of the $\Lambda$ resonances.  The
first state for which it is apparent~\cite{PDG} is the $\Lambda(1520)
\, D_{03}$ [note that the $\Lambda(1405) \, S_{01}$ lies below the
$\overline{K} N$ threshold].  This state is traditionally assigned to
be an SU(3) singlet.  The phase space ($\propto \! p^1$) for decay
into $\overline{K} N$ is only about 3/4 of that for $\pi \Lambda$, but
the BR for the former is actually slightly larger than for the latter.
Note however that the decay is a $d$ wave, so that the partial width
goes as the $p$ to the {\em fifth\/} power.  Thus the $\overline{K} N$
decay is kinematically suppressed by a factor of $\sim \! 4$--5
compared to the $\pi \Lambda$, so that the coupling is $\sim \! 4$--5
times larger.  The dominance of the $\overline{K} N$ coupling as what
one expects at large $N_c$ if the state is a singlet.

Virtually all of the low-lying $\Lambda$ resonances have substantial
$\overline{K} N$ BR, again suggesting a sizeable {\bf 1} component in
most $\Lambda$ resonances.  However, for most of these states the BR
are not determined with sufficient certainty to make definitive
statements.  Many of these states appear to have substantial BR to
both $\overline{K} N$ and $\pi \Lambda$.  To the extent that these
results are reliable, one has evidence for important effects of SU(3)
breaking, indicating to the mixing of SU(3) irreps.  The
$\Lambda(1830) \, D_{05}$ is a notable exception: Its BR to
$\overline{K} N$ is less than 10\%, strongly suggesting that it is
predominantly octet.

The situation with $\Sigma$ resonances is intriguing.  Since the only
SU(3) irreps available at $N_c \! = \! 3$ are {\bf 8} and {\bf 10},
the large $N_c$ selection rule suggests small $\overline{K} N$ BR\@.
While this is true for most of these resonances, a few [notably the
$\Sigma(1775) \, D_{15}$] have substantial $\overline{K} N$ couplings.
Nevertheless, such effects may well be $1/N_c$ corrections of a type
relatively easy to understand.  For any $N_c \! \ge \! 5$, the ``{\bf
S}'' irrep would contain $\Sigma$ resonances with large $\overline{K}
N$ couplings; in the final step of setting $N_c \!  = \! 3$, by
unitarity some part of these couplings must spill over into the SU(3)
{\bf 8} and {\bf 10} irreps.  It is very tempting to study these
$1/N_c$ effects simply by retaining the full arbitrary-$N_c$ CGC in
Eq.~(\ref{Mmaster}), but this is only one source of $1/N_c$
corrections; Eq.~(\ref{Mmaster}) in its current form only includes
amplitudes that survive the large $N_c$ limit.

Unfortunately, too little is known about $\Xi$ and $\Omega$
resonances~\cite{PDG} to perform an interesting analysis of this sort.

The next result of phenomenological interest to the $N_c \! = \! 3$
universe is that the $K \! = \! 0$ multiplet, (``{\bf 8}'', 1/2)
$\oplus$ (``{\bf 10}'', 3/2), couples to $\eta$ but not $\pi$, while
the other (``{\bf 8}'', 1/2) has $K \! = \! 1$ and couples to $\pi$
but not $\eta$.  More generally, the $K \! = \! 1$ pole appears only
in channels coupled to $\pi$.  These results are exact in the large
$N_c$ limit.  In the spin-3/2 case large $N_c$ provides ``{\bf 10}''s,
with $K \! = \!  0$, 1, and 2; however, for $N_c \! = \! 3$ only one
remains, and so in that case it is not obvious how to identify
physical states with the large $N_c$ multiplets.  Nevertheless, both
$N_c \! = \! 3$ and larger $N_c$ provide exactly 2 (``{\bf 8}'', 1/2)
multiplets, making the coupling prediction testable.  Indeed, the fact
that one of these physical resonances, $N(1535)$, is $\eta$-philic and
$\pi$-phobic, while the other, $N(1650)$, is the reverse, was the
original phenomenological evidence~\cite{CL1st} offered in support of
this type of analysis.

This effect appears in the state $\Lambda (1670) \, S_{01}$, which
lies a mere 5~MeV above the $\eta \Lambda$ threshold (a phase space
about 6 times smaller than that for $\pi \Sigma$), and yet has a BR to
this channel of 10--25\%.  This suggests that the state is
predominantly an $\eta$-philic $K \! = \! 0$ state.  Likewise, the
$\Sigma (1750) \, S_{11}$ lies only a few MeV above the $\eta \Sigma$
threshold but has a substantial (15--55\%) BR to that channel, and
therefore is also predominantly $K \! = \! 0$.  On the other hand,
$\Lambda (1800) \, S_{01}$ has no detected $\eta \Lambda$ coupling,
and therefore appears to be the $K \! = \! 1$ state.

One more interesting result of this analysis is a method of
distinguishing {\bf 8} and {\bf 10} resonances based upon their decay
modes.  One such category arises from SU(3) CGC that are smaller than
the saturation of the bound given in Eq.~(\ref{CGCmag}); in the cases
considered here, this occurs for $\eta \Sigma \to \Sigma$(``{\bf
10}''), for $\eta \Xi \to \Xi$(``{\bf 10}''), for $\eta \Sigma^* \to
\Sigma$(``{\bf 8}''), and for $\eta \Xi^* \to \Xi$(``{\bf 8}''), all
of which are $O(1/N_c)$ smaller than naively expected.  One then
concludes, for example, that a $\Sigma$ resonance with a large $\eta
\Sigma^*$ coupling (none such yet observed) is mostly {\bf 10}.
Another category arises from the interesting property that
Eq.~(\ref{Mmaster}) applied to $\pi \Lambda$ and $\pi \Sigma$ external
states differs in only by the isospin quantum number in the
external-state CGC\@.  In particular, using the CGC in
Ref.~\cite{CLSU3} one finds the amplitude ratios
\begin{eqnarray}
r_{\bf 8} & \equiv & \frac{{\cal A} [ \Sigma \pi \to \Sigma (\mbox{``{\bf
8}''}) ]}{{\cal A} [ \Sigma \pi \to \Lambda (\mbox{``{\bf 8}''}) ]} =
\frac{N_c(N_c+7)}{N_c+6} \sqrt{\frac{2}{(N_c+3)(N_c-1)}} \ , \nonumber \\
r_{\bf 10} & \equiv & \frac{{\cal A} [ \Sigma \pi \to \Sigma (\mbox{``{\bf
10}''}) ]}{{\cal A} [ \Sigma \pi \to \Lambda (\mbox{``{\bf 10}''}) ]} =
-\frac{N_c+1}{\sqrt{2(N_c+3)(N_c-1)}} \ .
\end{eqnarray}
The calculation of $r_{\bf 8}$ requires one to sum coherently over the
``{\bf 8}$_\gamma$'' irreps.  One finds $r_{\bf 8} (\infty) \! = \!
+\sqrt{2}$ and $r_{\bf 10} (\infty) \! = \! -1/\sqrt{2}$, which
explains why (as seen in Tables~\ref{t1}--\ref{t2}) scattering
amplitudes for $\Sigma$(``{\bf 8}'') prefer $\pi \Sigma$ to $\pi
\Lambda$ couplings by a 2:1 ratio, and those for
$\Sigma$(``{\bf 10}'') are the reverse.  The function $r_{\bf 8} (N_c)
/ r_{\bf 8} (\infty)$ equals $5/(3\sqrt{3}) \! \approx \! 0.96$ for
$N_c \! = \! 3$ and rises monotonically to 1 (for odd integers $N_c$)
as $N_c$ increases, while $r_{\bf 10} (N_c) / r_{\bf 10} (\infty)$
equals $2/\sqrt{3} \! \approx \! 1.15$ for $N_c \! = \!  3$ and drops
monotonically to 1 as $N_c$ increases.  While elucidating but one
source of $1/N_c$ breaking in the full amplitudes, this exercise gives
an indication of how well one might expect the large $N_c$ predictions
to work.

\section{Conclusions} \label{concl}

The $1/N_c$ expansion applied to the baryon resonance sector continues
to provide surprises, both in terms of the organization of states into
multiplets and the implications for couplings to asymptotic
meson-baryon states, which enter into production and decay processes.
We have shown that a remnant of the old quark-picture (``{\bf
70}''$\!$,$\, 1^-$) of SU(6)$\times$O(3) survives as a consequence of
the fundamental emergent SU(6) contracted spin-flavor symmetry at
large $N_c$: The (``{\bf 70}''$\!$,$\, 1^-$) is a reducible multiplet
whose remaining undetermined index, $K$, is the same one that
distinguishes the nonstrange multiplets.  In the 3-flavor case, $K$
assumes the 5 values $0, \frac 1 2 , 1, \frac 3 2, 2$, and distinct
SU(3) multiplets with the same $K$ value are degenerate in mass and
width up to $O(1/N_c)$ corrections.

We showed furthermore that both the SU(3) group theory and the
spin-flavor symmetry produce phenomenologically interesting
predictions that appear to be borne out where data is available.  The
former predicts that resonances in the {\bf 8} and {\bf 10}
representations of SU(3) prefer to decay via nonstrange mesons, while
those in the {\bf 1} prefer to decay via $\overline{K}$'s.  The latter
predicts that, of the two spin-$\frac 1 2$ {\bf 8}'s in (``{\bf
70}''$\!$,$\, 1^-$), one decays via $\eta$ and one via $\pi$.

Thus far, this analysis remains descriptive and exploratory.
Improvements require advances in both data measurement and and partial
wave analysis, as well as the theoretical method.  Anyone who has
examined the hyperon resonance section of the {\it Review of Particle
Physics}~\cite{PDG} will agree that improvements on published data and
methods of analysis will prove extremely useful in understanding the
physical baryon resonance sector.  From the theoretical point of view,
the most significant improvements required both fall into the category
of $1/N_c$ corrections.  For the 2-flavor system, it is known how to
incorporate $1/N_c$-suppressed amplitudes~\cite{CL1N}.  The
leading-order 2-flavor amplitudes in $1/N_c$ all assume an extremely
simple behavior when expressed in the $t$ channel: $I_t \! = \! J_t$,
while amplitudes with $|I_t \! - J_t| \! = \! n$ are suppressed by
$O(1/N_c^n)$.  The generalization of this rule to three flavors is one
of the next problems to tackle.  The other $1/N_c$ effect that must be
mastered is the nature of decoupling of the spurious states that only
occur for $N_c \! > \! 3$, such as isospin-$\frac 3 2$ $\Xi$'s.  Once
these effects are fully understood, the $1/N_c$ expansion will be
fully available to the 3-flavor baryon resonance sector in the same
way that chiral perturbation theory describes soft mesons.

{\it Acknowledgments.}  T.D.C.\ was supported by the D.O.E.\ through
grant DE-FGO2-93ER-40762; R.F.L.\ was supported by the N.S.F.\ through
grant PHY-0140362.

\end{document}